\documentstyle[prl,aps,epsf]{revtex}

\begin{document}
\def \beq {\begin{equation}}
\def \eeq {\end{equation}}
\def \bes {\begin{eqnarray}}
\def \ees {\end{eqnarray}}
\def \ni {\noindent}
\def \nn {\nonumber}
\def \z {\tilde{z}}
\def \kp {k_{\bot}}
\def \e {\varepsilon}
\def \dpa {\Delta_{\|}}
\def \dpp {\Delta_{\bot}}

\title{Surface-impedance approach solves problems with
the thermal Casimir force between real metals}

\author{
B.~Geyer,\footnote{E-mail: geyer@itp.uni-leipzig.de}
G.~L.~Klimchitskaya,\footnote{On leave from
North-West Polytechnical University,
St.Petersburg, Russia, and Federal University of 
Para\'{\i}ba,
Jo\~{a}o Pessoa, Brazil.
E-mail: galina@fisica.ufpb.br}
and V.~M.~Mostepanenko\footnote{On leave from
Noncommercial Partnership ``Scientific Instruments'',
Moscow, Russia, and Federal University of 
Para\'{\i}ba,
Jo\~{a}o Pessoa, Brazil.
E-mail: mostep@fisica.ufpb.br}
}

\address
{Center of Theoretical Studies and Institute for Theoretical Physics,\\
Leipzig University, Augustusplatz 10/11, 04109, Leipzig, Germany }

\maketitle
\begin{abstract}
 The surface impedance approach to the description of the thermal
Casimir effect in the case of real metals is elaborated starting
from the free energy of oscillators. The Lifshitz formula
expressed in terms of the dielectric permittivity
depending only on frequency
is shown to be
inapplicable in the frequency region where a real current may
arise leading to Joule heating of the metal. The
standard
concept of a
fluctuating electromagnetic field on such frequencies
meets difficulties when
used as a model for the zero-point oscillations or thermal photons
in the thermal equilibrium inside metals. Instead, the surface
impedance permits not to consider the electromagnetic oscillations
inside the metal but taking the realistic material properties into
account by means of the effective boundary condition. An
independent derivation of the Lifshitz-type formulas for the
Casimir free energy and force between two metal plates is
presented within the impedance approach. It is shown that they are
free of the contradictions with thermodynamics which are specific
to the usual Lifshitz formula for dielectrics in combination with
the Drude model. We demonstrate that in the impedance approach the
zero-frequency contribution is uniquely fixed by the form of
impedance function and does not need any of the {\it ad hoc}
prescriptions intensively discussed in the recent literature. As
an example, the computations of the Casimir free energy between
two gold plates (or the Casimir force acting between a plate and a
sphere) are performed at different separations and temperatures
specific for the regions of the anomalous skin effect and infrared
optics. The results are in good agreement with those obtained by
the use of the tabulated optical data for the complex refraction
index and plasma model. It is argued that the surface impedance
approach lays a reliable theoretical framework for the future
measurements of the thermal Casimir force.
\end{abstract}
\vskip 3mm
Pacs: 12.20.Ds, 42.50.Lc, 05.70.-a
\large
\section{Introduction}

Considerable attention has been focused recently on the Casimir
effect \cite{1} which is a rare manifestation of the zero-point
oscillations of the electromagnetic field at macroscopic scales.
The Casimir force arises as response to the change of the spectrum
of zero-point oscillations when material boundaries are present.
It acts between the boundaries of these bodies and depends on the
parameters of their materials, their geometry (including surface
roughness), and on the temperature (for a detailed information see
the monographs and reviews \cite{2,3,4,5,6}). Currently, many
precision measurements of the Casimir force between metal
boundaries have been performed \cite{7,8,9,10,11,12,13,14}. Their
results were used for constraining hypothetical forces predicted
by unified gauge theories of fundamental interactions
\cite{15,16,17} and in nanotechnological applications
\cite{18,19}.

With respect to the present state of the art, the theoretical
description of the Casimir force calls for a careful account of
all material properties and other relevant factors. Surprisingly,
it was found that the calculations of the temperature effect on
the Casimir force between real metals of finite conductivity run
into serious troubles which have been the subject of much
controversy \cite{20,21,22,23,24,25,26,27,28,29,30,31,32,33}. The
key contradiction is on whether the term of the Lifshitz formula
\cite{34,35} related to the zero Matsubara frequency for the {\em
perpendicular} polarized modes of an electromagnetic field
contributes to the physical quantities and, if so, how much would
its contribution be. At the moment there are five distinct
approaches to the resolution of this problem in the recent
literature, namely:
\begin{itemize}
\item[(a)]
According to the first approach, proposed in Ref.~\cite{20} and
supported in Refs.~\cite{26,27}, in the case of real metals the
term of the Lifshitz formula with zero Matsubara frequency should
be calculated by using the Drude dielectric function. As a result,
the perpendicular polarized modes do not contribute to this term.
\item[(b)]
In the other approach \cite{28} a special modification of the
zero-frequency term of the Lifshitz formula, supplemented by the
Drude model, was proposed leading to a nonzero contribution of the
perpendicular polarized modes. This modification was done by
analogy with the prescription of Ref.~\cite{36} for an ideal metal
but it does not coincide with it.
\item[(c)]
In the framework of the approaches of Refs.~\cite{23,24,25} the
modification of the zero-frequency term of the Lifshitz formula
was made identical to that for an ideal metal \cite{36}. As a
consequence, the contribution of the perpendicular polarized modes
to the zero-frequency term in Refs.~\cite{23,24,25} is nonzero and
coincides with that for an ideal metal.
\item[(d)]
According to Refs.~\cite{21,22,29} the contribution of the
perpendicular polarized modes with zero Matsubara frequency is
nonzero and should be calculated by substituting the free electron
plasma dielectric function into the unmodified Lifshitz formula.
\item[(e)]
Finally, according to the approach of Refs.~\cite{30,31}, the
description of the thermal Casimir force can be obtained by
using
the Leontovich surface impedance boundary condition (recently this
approach was applied also in Ref.~\cite{32}). In doing so, the
perpendicular polarized modes give a nonzero contribution to the
zero-frequency term of the Lifshitz formula prescribed by the form
of the impedance.
\end{itemize}
As to the contribution of the modes with a {\em parallel}
polarization to the zero-frequency term, there is consensus
between all these approaches that for real metals it is nonzero
and the same as for ideal metals. Note also that some of the
viewpoints varied with the time in the framework of the above
approaches (a) -- (e). For example, in Refs.~\cite{23,24} the
approach (c) was considered as an universal prescription, whereas
in Ref.~\cite{32} it is restricted by the range of only cryogenic
temperatures, and in Ref.~\cite{25} by the case of sufficiently
large separation distances and by the presence of thin covering
metallic films. The approach (b), proposed in Ref.~\cite{28} to
resolve the contradictions arising for the Drude model in
combination with the Lifshitz formula, was considered later in
Ref.~\cite{31} as unnecessary as the Drude model itself turned out
to be irrelevant for the description of the thermal Casimir force
between real metals. It should be particularly emphasized that the
approaches (a) and (c) were proved to be in contradiction
 with thermodynamics \cite{31,33} since they violate the Nernst
heat theorem. A detailed comparison of all the approaches can be
found in Refs.~\cite{28,31}.

The present paper aims to work out quite clearly that the surface
impedance approach provides an answer to all the complicated
problems with the retarded Casimir force between real metals at
both zero and nonzero temperature. We demonstrate that the main
reason why the Drude model in combination with the Lifshitz theory
had failed to describe the thermal Casimir force is the
inadequacy of the
standard
concept of a fluctuating electromagnetic field
depending only on frequency
inside a lossy real metal. Rather than to consider fluctuations
inside a metal, the surface impedance approach suggests that the
effective boundary conditions
take into consideration in a noncontradictory way
the involved reflection properties from
the surface of a real conductor. In this case no additional
prescriptions are needed, and the values of the zero-frequency
contributions for both parallel and perpendicular modes follow
immediately from the explicit form of the impedance function.

On this basis, we present for the first time a derivation of the
formula for the Casimir free energy and force in a configuration
of two parallel plates in the surface impedance approach (in
Ref.~\cite{30} it was presented without proof by the use of a
prescription changing the integration over continuous frequencies
for the summation over the discrete Matsubara frequencies). The
necessity of this derivation results from the fundamental role
played by the concept of the surface impedance in the theory of
the Casimir effect between real metals. The relationship between
this formula and the Lifshitz formula for the free energy is found
(the new formula is obtained by exchanging the reflection
coefficients, which appear in the Lifshitz formula, for the ones
derived in the surface impedance approach). Our derivation starts
from the free energy of the oscillators and suggests also other
means to derive the usual Lifshitz formula for the thermal Casimir
force between dielectrics. The obtained formula is applied to
compute the Casimir free energy and force at different
temperatures and separation distances between the test bodies. To
do this, one has to use the impedance functions describing the
regions of infrared optics, anomalous or normal skin effect
depending on the value of the characteristic frequency giving the
main contribution to the Casimir force. It is shown that no
contradictions with thermodynamics arise, and no artificial
prescriptions for the zero-frequency term are needed (we also
demonstrate that the predictions of large temperature corrections
to the Casimir force at small separations and cryogenic
temperatures, made in Ref.~\cite{32}, are in error since the
impedance of the anomalous skin effect was used in \cite{32}
outside its range of application).

The paper is organized as follows. In Sec.~II we demonstrate the
inadequacy of the concept of a fluctuating electromagnetic field
inside lossy real metals and remind the basic facts from the
theory of surface impedance. Sec.~III is devoted to the derivation
of the electromagnetic oscillation spectrum between two parallel
plates starting from the impedance boundary condition. In Sec.~IV
the formula for the Casimir free energy is derived in the surface
impedance approach. The new simple derivation of the usual
Lifshitz formula for the thermal Casimir force between dielectrics
is also given here. Sec.~V contains the calculations of the
Casimir energy and force at zero temperature using the impedance
approach. The results are compared with the previously known ones,
obtained by the use of the usual Lifshitz formula and the
tabulated optical data, and found to be in agreement. In Sec.~VI
the computations at nonzero temperature are performed in the
impedance approach at different separation distances between the
test bodies. They demonstrate good agreement in transition regions
between the different analytical expressions for the impedance. In
Sec.~VII the reader finds conclusions and a discussion of the
relationship between the proposed impedance approach to the
Casimir effect and the theory of the van der Waals forces valid at
small separation distances between the test bodies.

\section{The concept of a fluctuating electromagnetic field and
the surface impedance}

It is well known that the concept of a fluctuating electromagnetic
field works well for the description of zero-point oscillations
within media with a frequency-dependent dielectric permittivity
where no real electric current does arise. We will now look at a
conductor in an external electric field, which varies with some
frequency $\omega$ satisfying the conditions
    \beq
    l\ll\delta_n(\omega),\qquad l\ll\frac{v_F}{\omega},
\label{1}
    \eeq
\ni where $l$ is the mean free path of a conduction electron,
$\delta_n(\omega)=c/\sqrt{2\pi\sigma\omega}$ is the penetration
depth of the field inside a metal, $\sigma$ is the conductivity,
and $v_F$ is the Fermi velocity. Eqs.~(\ref{1}) determine the
domain of the normal skin effect \cite{37}. In this frequency
region the external field leads to the initiation of a real
current of the conduction electrons.

The normal skin effect is characterized by the volume relaxation
described by the temperature-dependent relaxation frequency
$\gamma(T)$. As a result, the mean free path of the conduction
electrons is also temperature-dependent, $l=l(T)=v_F/\gamma(T)$,
and increases with a decrease of temperature. The interaction of
the conduction electrons with the elementary excitations of the
crystal lattice (phonons) leads to the occurrence of electric
resistance and heating of the metal. The dielectric permittivity
of a metal in the domain of the normal skin effect can be modelled
by the Drude function
\beq
\e(\omega)=1-\frac{\omega_p^2}{\omega\left[\omega+i\gamma(T)\right]},
\label{2} \eeq \ni %
where $\omega_p$ is the plasma frequency of the free electron
plasma model ($\omega_p$ is temperature-independent). Remind that
the Drude dielectric function (\ref{2}) was used in the approaches
(a) -- (c) (see Introduction) in combination with the Lifshitz
formula to describe the thermal Casimir force between real metals
for the frequencies both inside and outside of the region
(\ref{1}). This has led to difficulties including the violation of
Nernst's heat theorem (see Introduction).

The physical reason for these difficulties becomes quite clear
when one observes that the
usual
alternating electric field with
frequencies characteristic for the normal skin effect inevitably
leads to heating of a metal as it penetrates through the skin
layer. By contrast, the thermal photons in thermal equilibrium
with a metal plate or, much less, the virtual photons (giving rise
to the Casimir effect) can not, under any circumstances, lead to
the initiation of a real current and heating of the metal (of
course, this is strictly prohibited by thermodynamics). Hence the
standard
concept of a fluctuating electromagnetic field penetrating inside
a metal
described by the Drude dielectric function fails to model
virtual and thermal photons in the
frequency region (\ref{1}). As a consequence, the Lifshitz formula
can not be applied in combination with the Drude dielectric
function (\ref{2}) to describe the thermal Casimir force even in
the domain of the normal skin effect.

These arguments are supported also by considering the other
frequency regions. At higher frequencies or larger $l$ (lower
temperatures) for most of the metals the anomalous skin effect
holds, which is characterized by the inequalities
\beq
\delta_a(\omega)\ll l,\qquad\delta_a(\omega)\ll\frac{v_F}{\omega},
\label{3} \eeq \ni%
where the skin depth is given by \cite{37}
\beq
\delta_a(\omega)=\left(\frac{4\pi c^2\hbar^3}{\omega
e^2S_F}\right)^{1/3},
\label{4} \eeq \ni%
and $S_F$ is the total area of the Fermi surface [in fact, within
the inequalities (\ref{3}) ``much less'' can be replaced by
``less'' thereby preserving Eq.~(\ref{4}) with a good precision].
In the frequency region of Eqs.~(\ref{3}) the volume relaxation is
not significant but the connection between the electric field and
current becomes nonlocal. Because of this, a metal can not be
described by any dielectric function depending only on the
frequency. As with the normal skin effect, this leads to the
inapplicability of the standard
concept of a fluctuating field
spreading
inside a medium
described by $\e(\omega)$ and to the impossibility to calculate
the Casimir force on this theoretical basis.
If one would like to preserve the role of a fluctuating field
within the domain of the anomalous skin effect, some nonlocal
generalization of this concept is required.
[Note that for some
metals, especially for alloys, instead of Eq.~(\ref{3}), the
inequalities $v_F/\omega\ll l\ll\delta$ hold, specifying the
so-called relaxation domain \cite{38,39}; here the space
dispersion is also essential.]

On further rise of the frequency, the inequalities hold
    \beq
    \frac{v_F}{\omega}\ll\delta_r\ll l,
    \label{5}
    \eeq \ni
where $\delta_r=c/\omega_p$, which determines the domain of
infrared optics (note that the condition $\hbar\omega\ll{\e}_F$ is
also supported where ${\e}_F=\hbar\omega_p$ is the Fermi energy).
In this domain the volume relaxation does not play any role and
the space dispersion is absent. Under the conditions (\ref{5})
metals can be described in the framework of the free-electron
plasma model with a frequency dependent dielectric permittivity
\beq \e(\omega)=1-\frac{\omega_p^2}{\omega^2}. \label{6} \eeq \ni
According to the plasma model, the conductivity is pure imaginary
and hence there is no any real current or heating due to an
electric field penetrating into the metal. Because of this, the
standard
concept of a fluctuating electromagnetic field, as a model for the
virtual and thermal photons, works well in the domain of infrared
optics. Remind that the plasma model dielectric permittivity in
combination with the Lifshitz formula was used to calculate the
thermal Casimir force [see approach (d) from the Introduction].
This approach did not meet any difficulties or contradictions with
the basic principles of thermodynamics and has led to physically
reasonable results. The domain of infrared optics is followed by
the domain of ultraviolet frequencies where metals become
transparent.

As is evident from the foregoing, the
standard
concept of a fluctuating
electromagnetic field, penetrating into a metal described by the
dielectric permittivity
depending only on frequency
can not be used as
an adequate
model for the virtual and thermal photons of some frequencies.
There is a frequency region (the domain of the normal skin effect)
where this model is in conflict with the basic properties of
virtual and thermal photons at equilibrium which, among other
things, can not lead to heating of a metal. In addition, in the
region of the anomalous skin effect and relaxation domain a metal
can not be described by the dielectric permittivity
depending only on frequency.
Because of this, another theoretical basis is
preferred
to find the thermal Casimir force between real metals different
from the one used in the case of dielectrics. Here we show that
this basis is given by the concept of the surface impedance
introduced by M.~A.~Leontovich \cite{35,40}.

The fundamental difference of the surface impedance approach from
the usual approaches is that it permits not to consider the
electromagnetic fluctuations inside a metal. Instead, the
appropriate boundary conditions are imposed taking into account
the properties of real metal
\beq
{\mbox{\boldmath{$E$}}}_t=Z(\omega)
\left[{\mbox{\boldmath{$B$}}}_t
\times {\mbox{\boldmath{$n$}}}\right],
\label{7}
\eeq \ni
where $Z(\omega)$ is the surface impedance of
the conductor, {\boldmath{$E$}}${}_t$ and {\boldmath{$B$}}${}_t$
are the tangential components of electric and magnetic fields, and
{\boldmath{$n$}} is the unit normal vector to the surface
(pointing inside the metal). The boundary condition (\ref{7}) can
be used to determine the electromagnetic field outside a metal.
Note, that the impedance $Z(\omega)$ and the condition (\ref{7})
suggest a more universal description than the one by means of
$\e$. They still hold under the inequalities (\ref{3}) where a
description in terms of the dielectric permittivity $\e(\omega)$
is impossible. For an ideal metal we have $Z\equiv 0$ and for real
nonmagnetic metals $|Z|\ll 1$ holds \cite{40}.

The calculation of the surface impedance over the whole frequency
axis is based on the kinetic theory \cite{35} and is rather
cumbersome. However, in the domains of the normal and the
anomalous skin effect, and also for infrared optics, simple
asymptotic expressions follow which are of great help to compute
the Casimir force between real metals. Thus, in the domain of the
normal skin effect, given by Eq.~(\ref{1}), the surface impedance
is \cite{35}
\beq
Z_n(\omega)=(1-i)\sqrt{\frac{\omega}{8\pi\sigma}}.
\label{8}
\eeq

In the domain of the anomalous skin effect, determined by
Eq.~(\ref{3}), the impedance depends on the shape of the Fermi
surface \cite{41}. For a polycrystalline metal, composed of many
single crystal grains of different orientations, one obtains an
approximately spherical Fermi surface and the impedance is
\cite{41} \beq Z_a(\omega)=\frac{2(1-i\sqrt{3})}{3\sqrt{3}}
\frac{\omega\delta_a(\omega)}{c}, \label{9} \eeq \ni where
$\delta_a(\omega)$ was defined in Eq.~(\ref{4}).

In the domain of infrared optics [see Eq.~(\ref{5})] the impedance
is given by \cite{35}
\beq
Z_r(\omega)=-i\frac{\omega}{\sqrt{\omega_p^2-\omega^2}}.
\label{10}
\eeq

Now one can impose the impedance boundary condition (\ref{7}) on
the surface of metal plates, find the oscillation spectrum in the
space between the plates and calculate the Casimir free energy
density and force without consideration of a fluctuating
electromagnetic field inside the metal. This was performed at zero
temperature in Ref.~\cite{42} (see also Ref.~\cite{3}). Another
approach, being similar in spirit, was used at $T=0$ in
Ref.~\cite{43} where the reflection coefficients in the Lifshitz
formula were expressed in terms of $Z(\omega)$.

Now, the question arises what expression for the impedance
(\ref{8}), (\ref{9}) or (\ref{10}) should be used to calculate the
Casimir effect. To answer this question, it is good to bear in the
mind that the main contribution to the Casimir free energy and
force is given by the frequency region centered around the
so-called characteristic frequency $\omega_c=c/(2a)$, where $a$ is
the space separation between the two bodies, parallel plates for
instance. The value of $\omega_c$ may belong to the frequency
region given by Eqs.~(\ref{1}), (\ref{3}) or (\ref{5}), with the
result that the functions  (\ref{8}), (\ref{9}) or (\ref{10}),
respectively, should be used defining the impedance in the domains
of the normal or anomalous skin effect and infrared optics.

By way of example, for most of the metals ($Au$ for instance) at
room temperature the application region (\ref{1}) of the normal
skin effect with the impedance (\ref{8}) extends up to the
frequencies of order $10^{12}\,$rad/s. The application region
(\ref{3}) of the anomalous skin effect at $T=300\,$K is very
narrow and extends up to around $(6-7)\times 10^{13}\,$rad/s. It
should be stressed, however, that with the decrease of temperature
the application region of the normal skin effect practically
disappears and the anomalous skin effect extends to all
frequencies being lesser than $10^{12}\,$rad/s. The reason is that
$l$ increases and $\delta_n(\omega)$ decreases with the decrease
of temperature. As a result, the first inequality in Eq.~(\ref{1})
breaks down, whereas the first inequality in Eq.~(\ref{3}) is
satisfied also at smaller frequencies. Finally, the impedance of
the infrared optics (\ref{10}) is applicable up to the frequencies
of order $0.1\omega_p$ (for $Au$, for instance,
$\omega_p=1.37\times 10^{16}\,$rad/s). It should be particularly
emphasized that the transition frequency between the anomalous
skin effect and infrared optics does not depend on the temperature
because all the parameters in the second inequality of Eq.~(\ref{3})
are temperature independent (at temperatures much smaller than the
Fermi temperature, which is of order $10^5\,$K). In Secs.~V and VI
we will discuss with more details which impedance function should be
used for the calculation of the Casimir force at different
separation distances between the test bodies (also the transition
regions between different impedance functions will be considered).

\section{Electromagnetic oscillation spectrum between two plates
in the surface impedance approach}

Here the derivation of the photon eigenfrequencies in the
framework of the surface impedance approach is presented. They are
needed to derive the Lifshitz-type formula for the free energy in
the case of real metals.

We consider the configuration of two parallel uncharged metal
plates, separated by a distance $a$, at a temperature $T$ in
thermal equilibrium. Let their nearest boundary planes be
described by the equations $z=\pm a/2$. We impose the boundary
condition (\ref{7}) on planes $z=\pm a/2$ and determine the
eigenfrequencies of the electromagnetic field in the free space
between the plates. The solutions of the Maxwell equations in
vacuum can be found in the form
\bes &&
{\mbox{\boldmath{$E$}}}_{\alpha}
(t,{\mbox{\boldmath{$r$}}})={\mbox{\boldmath{$e$}}}_p (k_{\bot},z)
\exp({i{\mbox{{\boldmath{$k_{\!\bot}r_{\!\bot}$}}}}-i\omega t}),
\nn \\
&& {\mbox{\boldmath{$B$}}}_{\alpha}
(t,{\mbox{\boldmath{$r$}}})={\mbox{\boldmath{$g$}}}_p (k_{\bot},z)
\exp({i{\mbox{{\boldmath{$k_{\!\bot}r_{\!\bot}$}}}}-i\omega t}).
\label{11} \ees \ni Here
{\boldmath{$r$}}$=(x,y,z)=({\mbox{\boldmath{$r$}}}_{\!\bot},z)$,
$\alpha=\{p,{\mbox{\boldmath{$k$}}}_{\!\bot},\omega\}$,
${\mbox{\boldmath{$k$}}}_{\!\bot}=(k_1,k_2)$ is the wave vector in
the plane $(x,\,y)$, $\omega$ is a frequency, and the index
$p=\|,\,\bot$ labels the two independent polarization states ($\|$
stands for the electric field parallel to the plane formed by
${\mbox{\boldmath{$k$}}}_{\!\bot}$ and $z$ axis, and $\bot$ stands
for the electric field perpendicular to this plane). From the
Maxwell equations the oscillatory equations for the functions
{\boldmath{$e$}}${}_p$, {\boldmath{$g$}}${}_p$ follow,
 \bes &&
\mbox{{\boldmath{$e$}}}_p^{\prime\prime}(k_{\!\bot},z)-
q^2(k_{\!\bot},\omega)\mbox{{\boldmath{$e$}}}_p(k_{\!\bot},z)=0,
\nn \\
&& \mbox{{\boldmath{$g$}}}_p^{\prime\prime}(k_{\!\bot},z)-
q^2(k_{\!\bot},\omega)\mbox{{\boldmath{$g$}}}_p(k_{\!\bot},z)=0,
\label{12} \ees \ni where $q^2(k_{\!\bot},\omega)\equiv q^2\equiv
k_{\!\bot}^2-\omega^2/c^2$, the prime denotes the derivative with
respect to $z$, and also the first-order equations
    \bes &&
e_{p,3}^\prime(k_{\!\bot},z) +ik_1e_{p,1}(k_{\!\bot},z)+
ik_2e_{p,2}(k_{\!\bot},z)=0,
\nn \\
&& g_{p,3}^\prime(k_{\!\bot},z) +ik_1g_{p,1}(k_{\!\bot},z)+
ik_2g_{p,2}(k_{\!\bot},z)=0 \label{13} \ees \ni (lower indices
1,\,2,\,3 after a comma stand for the projections of vectors
{\boldmath{$e$}}${}_p$, {\boldmath{$g$}}${}_p$ onto the axes
$x,\,y,\,z$, respectively).

Substituting Eqs.~(\ref{11}) into the boundary condition
(\ref{7}), using the Maxwell equations to express the magnetic field
and taking the direction of the normal into account
[{\boldmath{$n$}}=$(0,\,0,\,1)$ at the plane $z=a/2$, and
{\boldmath{$n$}}=$(0,\,0,\,-1)$ at $z=-a/2$], we find at the
boundaries $z=\pm a/2$, respectively,
    \bes &&
e_{p,1}\left(k_{\!\bot},\pm\frac{a}{2}\right)=
\pm\frac{iZc}{\omega}\left[ik_1\,
e_{p,3}\left(k_{\!\bot},\pm\frac{a}{2}\right)-
e_{p,1}^{\prime}\left(k_{\!\bot},\pm\frac{a}{2}\right)\right],
\nn \\
&&
e_{p,2}\left(k_{\!\bot},\pm\frac{a}{2}\right)=
\pm\frac{iZc}{\omega}\left[ik_2\,
e_{p,3}\left(k_{\!\bot},\pm\frac{a}{2}\right)-
e_{p,2}^{\prime}\left(k_{\!\bot},\pm\frac{a}{2}\right)\right].
\label{14}
\ees
\ni
The same boundary conditions for {\boldmath{$g$}}${}_p$ can be
obtained also.

Now, let us consider separately the cases of parallel and
perpendicular polarizations beginning with the parallel one.
Without loss of generality, we temporarily assume that $k_2=0$. In
this case $e_{\|,2}(k_{\!\bot},z)\equiv 0$ and the solution of
Eq.~(\ref{12}) has the form \beq e_{\|,1}(k_{\!\bot},z)=B\sinh
qz,\qquad e_{\|,3}(k_{\!\bot},z)=B\cosh qz, \label{15} \eeq \ni
where $A$ and $B$ do not depend on $z$. From Eq.~(\ref{13}) it
follows $Aq+ik_1B=0$. For the sake of convenience, we choose \beq
A=-\frac{ik_1}{q}\,e^{-aq/2}, \qquad B=e^{-aq/2}. \label{16} \eeq
\ni Substituting Eqs.~(\ref{15}) and (\ref{16}) into the impedance
boundary condition (\ref{14}), one obtains the dispersion equation
for the spectrum of the electromagnetic oscillations between
plates: \beq \Delta_{\|}^{\!(1)}(\omega,k_{\!\bot})\equiv
e^{-aq/2} \left(\sinh\frac{aq}{2}-\frac{iZ\omega}{cq}
\cosh\frac{aq}{2}\right)=0. \label{17} \eeq \ni Eqs.~(\ref{12}),
(\ref{13}) also have the solutions \bes &&
e_{\|,1}(k_{\!\bot},z)=e^{-aq/2}\cosh qz, \qquad
e_{\|,2}(k_{\!\bot},z)=0,
\nn \\
&& e_{\|,3}(k_{\!\bot},z)=-\frac{ik_1}{q}e^{-aq/2}\sinh qz.
\label{18} \ees \ni After substitution of Eq.~(\ref{18}) into
Eq.~(\ref{14}) a further dispersion equation for the modes with
parallel polarization is obtained: \beq
\Delta_{\|}^{\!(2)}(\omega,k_{\!\bot})\equiv e^{-aq/2}
\left(\cosh\frac{aq}{2}-\frac{iZ\omega}{cq}
\sinh\frac{aq}{2}\right)=0. \label{19} \eeq \ni It is obvious that
in Eqs.~(\ref{17}) and (\ref{19})
{\boldmath{$k$}}${}_{\!\bot}=(k_1,k_2)$ can be now considered as
arbitrary.

Exactly the same procedure is applicable to the case of the
perpendicular polarization. Once again, assuming  temporarily
$k_2=0$, we obtain the solutions of Eqs.~(\ref{12}) and (\ref{13})
in the form 
\beq
e_{\!\bot,1}(k_{\!\bot},z)=e_{\!\bot,3}(k_{\!\bot},z)=0 \label{20}
\eeq 
\ni and 
\beq e_{\!\bot,2}(k_{\!\bot},z)=e^{-aq/2}\sinh qz
\quad \mbox {or}
\quad
 e_{\!\bot,2}(k_{\!\bot},z)=e^{-aq/2}\cosh qz. 
\label{21} \eeq
\ni Substituting these solutions into Eq.~(\ref{14}), we arrive at
two dispersion equations for the determination of the
electromagnetic eigenfrequencies with perpendicular polarization
\bes && \Delta_{\bot}^{\!(1)}(\omega,k_{\!\bot})\equiv e^{-aq/2}
\left(\sinh\frac{aq}{2}+\frac{iZcq}{\omega}
\cosh\frac{aq}{2}\right)=0,
\nn \\
&& \Delta_{\bot}^{\!(2)}(\omega,k_{\!\bot})\equiv e^{-aq/2}
\left(\cosh\frac{aq}{2}+\frac{iZcq}{\omega}
\sinh\frac{aq}{2}\right)=0. \label{22} \ees \ni
Let us denote the
solutions of the transcendental equations (\ref{17}), (\ref{19})
by $\omega_{k_{\bot},n}^{\|}$, and the solutions of the
transcendental equations (\ref{22})  by
$\omega_{k_{\bot},n}^{\bot}$. Multiplying Eqs.~(\ref{17}),
(\ref{19}), we can finally find $\omega_{k_{\bot},n}^{\|}$ from
the equation \bes && \Delta_{\|}(\omega,k_{\!\bot})\equiv
\Delta_{\|}^{\!(1)}(\omega,k_{\!\bot})\,
\Delta_{\|}^{\!(2)}(\omega,k_{\!\bot})
\label{23} \\
&&\phantom{aa}=
\frac{1}{2}e^{-aq}\left(1-\eta^2\right)\left(\sinh aq-
\frac{2i\eta}{1-\eta^2}\cosh aq\right)=0,
\nn
\ees
\ni
where $\eta=\eta(\omega)=Z\omega/(cq)$.

In perfect analogy to this, by multiplication of Eqs.~(\ref{22}),
one can find the equation for the determination of
$\omega_{k_{\bot},n}^{\bot}$: \bes &&
\Delta_{\bot}(\omega,k_{\!\bot})\equiv
\Delta_{\bot}^{\!(1)}(\omega,k_{\!\bot})\,
\Delta_{\bot}^{\!(2)}(\omega,k_{\!\bot})
\label{24} \\
&&\phantom{aa}=
\frac{1}{2}e^{-aq}\left(1-\kappa^2\right)\left(\sinh aq+
\frac{2i\kappa}{1-\kappa^2}\cosh aq\right)=0,
\nn
\ees
\ni
where $\kappa=\kappa(\omega)=Zcq/\omega$.

Note that we have obtained the conditions for the determination
of the electromagnetic oscillation spectrum by the use of equations
for {\boldmath{$e$}}${}_p$. Exactly the same spectrum is obtained
if the equations for {\boldmath{$g$}}${}_p$ are used.

\section{Casimir free energy in 
the surface impedance approach}

Now we are in a position to present a rigorous derivation of
Lifshitz-type formulas for the Casimir free energy and force for
the configuration of two plates at a temperature $T$ in thermal
equilibrium in the surface impedance approach. As shown below,
these formulas are well adapted for the calculation of the Casimir
effect between real metals and are not subject to the
disadvantages of the approaches (a) -- (d) discussed in the
Introduction.

First we consider the case of real eigenfrequencies
$\omega_{k_{\bot},n}^{\|}$, $\omega_{k_{\bot},n}^{\bot}$ (this is
fulfilled for the pure imaginary impedance). The total free energy
of the electromagnetic oscillations is given by the sum of the
free energies of separate oscillators over all possible values of
their quantum numbers, \beq {\cal{F}}= \sum\limits_{\alpha}\left[
\frac{\hbar\omega_{\alpha}}{2}+k_BT\ln
\left(1-e^{-\frac{\hbar\omega_{\alpha}}{k_BT}}\right)\right],
\label{25} \eeq \ni where $k_B$ is the Boltzmann constant.
Identically, Eq.~(\ref{25}) can be rewritten as \beq
{\cal{F}}=k_BT \sum\limits_{\alpha}
\ln\left(2\sinh{\frac{\hbar\omega_{\alpha}}{2k_BT}}\right).
\label{26} \eeq \ni It is clear that at $T\to 0$, the value of
$\cal{F}$ from Eqs.~(\ref{25}), (\ref{26}) coincides with the sum
of the zero-point energies which is the traditional starting point
in theoretical investigations of the Casimir effect at zero
temperature.

Applying this to the electromagnetic oscillations between metal
plates, where $\alpha=\{p,{\mbox{$k$}}_{\!\bot},n\}$,
we obtain
    \beq {\cal{F}}=k_BT
\int_0^{\infty}\frac{k_{\!\bot}\,dk_{\!\bot}}{2\pi}
\sum\limits_{n}\left[
\ln\left(2\sinh{\frac{\hbar\omega_{k_{\bot},n}^{\|}}{2k_BT}}\right)
+\ln\left(2\sinh{\frac{\hbar\omega_{k_{\bot},n}^{\bot}}{2k_BT}}\right)
    \right]. \label{27} \eeq \ni
According to the calculations of Sec.~III, the eigenfrequencies of
the electromagnetic field between plates with parallel and
perpendicular polarizations are determined by Eqs.~(\ref{23}) and
(\ref{24}), respectively.

The expression in the right-hand side of Eq.~(\ref{27}) is
evidently infinite. Before performing a renormalization, let us
equivalently represent the sum over the eigenfrequencies
$\omega_{k_{\bot},n}^{\|,\bot}$ by the use of the argument theorem
like it is usually done in the derivation of the Lifshitz formula
at zero temperature by the method of surface modes \cite{6,44,45}.
Then Eq.~(\ref{27}) can be rewritten as
    \beq
{\cal{F}}=k_BT
\int_0^{\infty}\frac{k_{\!\bot}\,dk_{\!\bot}}{2\pi} \frac{1}{2\pi
i}\oint_{C_1} \ln\left(2\sinh{\frac{\hbar\omega}{2k_BT}}\right)
d\left[\ln\Delta_{\|}(\omega,k_{\!\bot})+
\ln\Delta_{\bot}(\omega,k_{\!\bot})\right].
\label{28}
\eeq
    \ni
Here, the closed contour $C_1$ is bypassed counterclockwise. It
consists of two arcs, one having an infinitely small radius
$\varepsilon$ and the other one an infinitely large radius $R$,
and two straight lines $L_1,\,L_2$ inclined at the angles $\pm 45$
degrees to the real axis (see Fig.~1a). The quantities
$\Delta_{\|,\bot}(\omega,k_{\!\bot})$, having their roots at the
photon eigenfrequencies, are defined in Eqs.~(\ref{23}),
(\ref{24}). Note that, unlike the usual derivation of the Lifshitz
formula at nonzero temperature \cite{2}, the function under the
integral in (\ref{28}) has branch points rather than poles at the
imaginary frequencies $\omega_l=i\xi_l$, where
    \beq
\xi_l=\frac{2\pi k_BTl}{\hbar}, \quad l=0,\,\pm 1,\,\pm 2,\,\ldots
\label{29}
    \eeq \ni
are the Matsubara frequencies. The contour $C_1$ in Fig.~1a is
chosen so as to avoid all these branch points and to enclose all the
photon eigenfrequencies.

The integral in Eq.~(\ref{28}) can be calculated as follows:
     \bes
&& I_{\|,\bot}\equiv \frac{1}{2\pi i}\oint_{C_1}
\ln\left(2\sinh{\frac{\hbar\omega}{2k_BT}}\right)
d\ln\Delta_{\|,\bot}(\omega,k_{\!\bot})
\label{30} \\
&&\phantom{aa}
=\frac{1}{2\pi i}\left[
\int_{L_2}\ln\left(2\sinh{\frac{\hbar\omega}{2k_BT}}\right)
d\ln\Delta_{\|,\bot}(\omega,k_{\!\bot})+
\int_{C_R}\ln\left(2\sinh{\frac{\hbar\omega}{2k_BT}}\right)
d\ln\Delta_{\|,\bot}(\omega,k_{\!\bot})\right.
\nn \\
&&\phantom{aa}
\left. +
\int_{L_1}\ln\left(2\sinh{\frac{\hbar\omega}{2k_BT}}\right)
d\ln\Delta_{\|,\bot}(\omega,k_{\!\bot})+
\int_{C_{\varepsilon}}\ln\left(2\sinh{\frac{\hbar\omega}{2k_BT}}\right)
d\ln\Delta_{\|,\bot}(\omega,k_{\!\bot})\right].
\nn
    \ees\ni
The integral along the arc of infinitely large radius $C_R$ vanishes
which follows from Eqs.~(\ref{23}), (\ref{24}) under the natural
conditions
\beq
\lim\limits_{\omega\to\infty}Z(\omega)=\mbox{const}, \qquad
\lim\limits_{\omega\to\infty}\frac{dZ(\omega)}{d\omega}=0.
\label{31}
\eeq

Integrating by parts in the right-hand side of Eq.~(\ref{30}), one obtains
\bes
&&
I_{\|,\bot}=
\frac{1}{2\pi i}\left[
\ln\left(2\sinh{\frac{\hbar\omega}{2k_BT}}\right)
\ln\Delta_{\|,\bot}(\omega,k_{\!\bot})
\biggl|_{-i\varepsilon}^{A}-\frac{\hbar}{k_BT}
\int_{L_2}\coth{\frac{\hbar\omega}{2k_BT}}
\ln\Delta_{\|,\bot}(\omega,k_{\!\bot})d\omega
\right.
\nn \\
&&
\phantom{aa}
+\ln\left(2\sinh{\frac{\hbar\omega}{2k_BT}}\right)
\ln\Delta_{\|,\bot}(\omega,k_{\!\bot})
\biggl|_{B}^{i\varepsilon}-\frac{\hbar}{k_BT}
\int_{L_1}\coth{\frac{\hbar\omega}{2k_BT}}
\ln\Delta_{\|,\bot}(\omega,k_{\!\bot})d\omega
\label{32}
\\
&& \phantom{aa} \left.
+\ln\left(2\sinh{\frac{\hbar\omega}{2k_BT}}\right)
\ln\Delta_{\|,\bot}(\omega,k_{\!\bot})
\biggl|_{i\varepsilon}^{-i\varepsilon}-\frac{\hbar}{k_BT}
\int_{C_{\varepsilon}}\coth{\frac{\hbar\omega}{2k_BT}}
\ln\Delta_{\|,\bot}(\omega,k_{\!\bot})d\omega\right], \nn
    \ees \ni
where the contours $L_{1,2}$ and the points $A,\,B$ are shown in
Fig.~1a. It is evident that all terms, besides the integrals,
cancel each other or are equal to zero (at the points $A,\,B$).
The integral along the $L_1$ can be calculated by the application
of the Cauchy theorem to the closed contour $C_2$ (see Fig.~1b)
inside of which the function under consideration is analytic
    \beq
-\int_{L_1}\coth{\frac{\hbar\omega}{2k_BT}}
\ln\Delta_{\|,\bot}(\omega,k_{\!\bot})d\omega =
-\int_{i\infty}^{i\varepsilon}\coth{\frac{\hbar\omega}{2k_BT}}
\ln\Delta_{\|,\bot}(\omega,k_{\!\bot})d\omega. \label{33}
    \eeq \ni
Here it is taken into account that the integral along $C_R$
vanishes. The path $(i\infty;i\varepsilon)$ contains semicircles
of radius $\varepsilon$ about the singular points $i\xi_l$ (poles)
of the function $\coth(\hbar\omega/2k_BT)$. The analogous formula
for the integral along the line $L_2$ is
    \beq
-\int_{L_2}\coth{\frac{\hbar\omega}{2k_BT}}
\ln\Delta_{\|,\bot}(\omega,k_{\!\bot})d\omega =
-\int_{-i\varepsilon}^{-i\infty}\coth{\frac{\hbar\omega}{2k_BT}}
\ln\Delta_{\|,\bot}(\omega,k_{\!\bot})d\omega.
    \label{34} \eeq \ni
Substituting Eqs.~(\ref{33}), (\ref{34}) into Eq.~(\ref{32}), one
arrives at
    \beq I_{\|,\bot}=-\frac{\hbar}{2\pi ik_BT}
\int_{i\infty}^{-i\infty}\coth{\frac{\hbar\omega}{2k_BT}}
\ln\Delta_{\|,\bot}(\omega,k_{\!\bot})d\omega.
    \label{35} \eeq \ni
The integration in Eq.~(\ref{35}) involving poles at the points
$i\xi_l$ leads to
    \bes
    && I_{\|,\bot}=\frac{i\hbar}{2\pi k_BT}
\int_{\infty}^{-\infty}\cot{\frac{\hbar\xi}{2k_BT}}
\ln\Delta_{\|,\bot}(\xi,k_{\!\bot})d\xi
\label{36} \\
&& \phantom{aaa} -\pi \sum\limits_{l=-\infty}^{\infty}
{\mbox{res}}
\left[ \coth{\frac{\hbar\omega}{2k_BT}}
\ln\Delta_{\|,\bot}(\omega,k_{\!\bot});i\xi_l\right], \nn \ees
    \ni
where the functions $\Delta_{\|,\bot}(\xi,k_{\!\bot})$ are
obtained from $\Delta_{\|,\bot}(\omega,k_{\!\bot})$ by the
substitution $\omega=i\xi$. In the case of real eigenfrequencies,
which is under consideration now, $\Delta_{\|,\bot}$ are even
functions of $\omega$ (and $\xi$). As a consequence, the seemingly
pure imaginary integral in the right-hand side of Eq.~(\ref{36})
vanishes. After the calculation of the residues, and using the
evenness of the functions $\Delta_{\|,\bot}(\omega,k_{\!\bot})$,
the result is
    \beq
I_{\|,\bot}=\sum\limits_{l=0}^{\infty}{\vphantom{\sum}}^{\prime}
\ln\Delta_{\|,\bot}(\xi_l,k_{\!\bot}), \label{37} \eeq \ni where
the prime on the summation sign means that the term for $l=0$ has
to be multiplied by 1/2.

Substituting the values (\ref{37}) of the integrals (\ref{30})
into Eq.~(\ref{28}), we find the equivalent but more simple
expression for the Casimir free energy
    \beq
{\cal{F}}=\frac{k_BT}{2\pi}
\int_0^{\infty}{k_{\!\bot}\,dk_{\!\bot}}
\sum\limits_{l=0}^{\infty}{\vphantom{\sum}}^{\prime}\left[
\ln\Delta_{\|}(\xi_l,k_{\!\bot})+
\ln\Delta_{\bot}(\xi_l,k_{\!\bot})\right].
\label{38}
\eeq

Expression (\ref{38}) is still infinite. To remove the
divergences, we subtract from the right-hand side of
Eq.~(\ref{38}) the free energy in the case of infinitely separated
interacting bodies ($a\to\infty$). Then the physical,
renormalized, free energy vanishes for infinitely remote plates.
{}From Eqs.~(\ref{23}), (\ref{24}) after the substitution $\omega\to
i\xi_l$ in the limit $a\to\infty$ it follows
    \bes &&
\Delta_{\|}^{\!\infty}(\xi_l,k_{\!\bot})=\frac{1}{4}\left(1+\eta_l^2\right)
\left(1+\frac{2\eta_l}{1+\eta_l^2}\right),
\nn \\
&&
\Delta_{\bot}^{\!\infty}(\xi_l,k_{\!\bot})=\frac{1}{4}
\left(1+\kappa_l^2\right)
\left(1+\frac{2\kappa_l}{1+\kappa_l^2}\right).
\label{39}
\ees
    \ni
The renormalization prescription is equivalent to the change of
$\Delta_{\|,\bot}(\xi_l,k_{\!\bot})$ in Eq.~(\ref{38}) for
    \beq
\Delta_{\|,\bot}^{\! R}(\xi_l,k_{\!\bot})\equiv
\frac{\Delta_{\|,\bot}(\xi_l,k_{\!\bot})}{\Delta_{\|,\bot}^{\!\infty}
(\xi_l,k_{\!\bot})}=
1-r_{\|,\bot}^{2}(\xi_l,k_{\!\bot})e^{-2aq_l},
\label{40}
\eeq
    \ni
where the quantities $r_{\|,\bot}(\xi_l,k_{\!\bot})$ have the
meaning of the reflection coefficients and are given by
    \bes
&&
r_{\|}^{2}(\xi_l,k_{\!\bot})=
\left(\frac{1-\eta_l}{1+\eta_l}\right)^2=
\left(\frac{cq_l-Z_l\xi_l}{cq_l+Z_l\xi_l}\right)^2,
\nn \\
&&
r_{\bot}^{2}(\xi_l,k_{\!\bot})=
\left(\frac{1-\kappa_l}{1+\kappa_l}\right)^2=
\left(\frac{\xi_l-Z_lcq_l}{\xi_l+Z_lcq_l}\right)^2.
\label{41}
\ees
    \ni
Here $Z_l\equiv Z(i\xi_l)$ and $q_l^2=k_{\!\bot}^2+\xi_l^2/c^2$.
The reflection coefficients (\ref{41}) are in accordance with
Ref.~\cite{40} where the reflection of a plane electromagnetic
wave incident from vacuum onto the plane surface of the metal was
described in terms of the surface impedance.

In such a manner the final renormalized expression for the Casimir
free energy in the surface impedance approach is given by
    \beq
{\cal{F}_R}=\frac{k_BT}{2\pi}
\int_0^{\infty}{k_{\!\bot}\,dk_{\!\bot}}
\sum\limits_{l=0}^{\infty}{\vphantom{\sum}}^{\prime}\left\{
\ln\left[1-r_{\|}^{2}(\xi_l,k_{\!\bot})e^{-2aq_l}\right] \right.
+\left.
\ln\left[1-r_{\bot}^{2}(\xi_l,k_{\!\bot})e^{-2aq_l}\right]
\right\},
\label{42}
\eeq
\ni
where the reflection coefficients are given by Eq.~(\ref{41}).

The Casimir force, acting between plates, is obtained from Eq.~(\ref{42})
\bes
&&
F=-\frac{\partial{\cal{F}}_R}{\partial a}=-\frac{k_BT}{\pi}
\int_0^{\infty}{k_{\!\bot}\,dk_{\!\bot}}
\sum\limits_{l=0}^{\infty}{\vphantom{\sum}}^{\prime}q_l
\label{43} \\
&&
\phantom{a}
\times
\left\{\left[r_{\|}^{-2}(\xi_l,k_{\!\bot})e^{2aq_l}-1\right]^{-1}+
\left[r_{\bot}^{-2}(\xi_l,k_{\!\bot})e^{2aq_l}-1\right]^{-1}
\right\}.
\nn
\ees

The above derivation was performed under the assumption that the
photon eigenfrequencies are real. This is, however, not the case
for arbitrary complex impedance. If the photon eigenfrequencies
are complex, the free energy is not given by Eq.~(\ref{26}) (which
is already clear from the complexity of the right-hand side of
this equation). For arbitrary complex impedance the correct
expression for the free energy should be determined from the
solution of an auxiliary electrodynamic problem \cite{46}. It
turns out that the Casimir free energy and force are the
functionals of the impedance even when the impedance has a nonzero
real part taking absorption into account. The solution of the
auxiliary electrodynamic problem leads to conclusion \cite{46}
that the correct free energy is obtained from
Eqs.~(\ref{38})--(\ref{42}) by analytic continuation to arbitrary
complex impedance, i.e., to arbitrary oscillation spectra. The
qualitative reason for the validity of this statement is that the
free energy depends only on the behavior of $Z(\omega)$ at the
imaginary frequency axis where $Z(\omega)$ is always real [see,
e.g., Eqs.~(\ref{8}) -- (\ref{10})]. Note that in the case of
complex eigenfrequencies exactly Eqs.~(\ref{42}), (\ref{43})
should be used written in terms of summations from zero to
infinity. Although for real eigenfrequencies the summations over
$l$ from $-\infty$ to $\infty$ can be equivalently used it is not
so for complex $\omega_{\alpha}$ as the dispersion functions
$\Delta_{\|,\bot}$ cease to be even any more \cite{46}.

It is necessary to stress that the above derivation of the free
energy in the impedance approach can be simply modified in order
to present the new derivation of the usual Lifshitz formula
describing the thermal Casimir force between dielectrics. In fact,
nothing should be changed in the presentation of this section
except for the explicit expressions of the dispersion functions
$\Delta_{\|,\bot}$ in Eq.~(\ref{38}) and thus of the reflection
coefficients $r_{\|,\bot}$ in Eqs.~(\ref{42}) and (\ref{43}). The
dispersion functions should be determined not according to
Sec.~III but from the consideration of a fluctuating
electromagnetic field both inside and outside of the dielectric
plates with the usual boundary conditions at the interfaces
\cite{2,6,44,45}. As a result, the Lifshitz reflection
coefficients take the form
    \beq
 r_{\|,L}^{2}(\xi_l,k_{\!\bot})=
\left(\frac{\varepsilon_lq_l-k_l}{\varepsilon_lq_l+k_l}\right)^2,
\quad 
r_{\bot,L}^{2}(\xi_l,k_{\!\bot})=
\left(\frac{q_l-k_l}{q_l+k_l}\right)^2,
\label{44}
\eeq \ni
where
$\varepsilon_l\equiv\varepsilon(i\xi_l)$, $\varepsilon(\omega)$ is
the dielectric permittivity of the plate material, and
$k_l^2\equiv k_{\!\bot}^2+\varepsilon_l\xi_l^2/c^2$. Then the
Lifshitz expressions for the Casimir free energy and force between
dielectrics are given by Eqs.~(\ref{42}), (\ref{43}) where the
substitution $r_{\|},\,r_{\bot}\to r_{\|,L},\,r_{\bot,L}$ is made.
In such a manner, we have performed also a new derivation of the
usual Lifshitz formula between dielectric plates starting from the
free energy of an oscillator. Conversely, the free energy and
force in the framework of the impedance approach are obtained from
the Lifshitz formula if the Fresnel-type reflection coefficients
$r_{\|,L},\,r_{\bot,L}$ are changed for those obtained by the use
of the impedance boundary condition.

It should be stressed, however, that the reflection coefficients
(\ref{44}) differ essentially from the impedance coefficients
(\ref{41}). To take an example, it is not possible to obtain the
coefficients (\ref{41}) from (\ref{44}) even if both descriptions
in terms of $\varepsilon(\omega)$ and $Z(\omega)$ are applicable
and the impedance is expressed in terms of the dielectric
permittivity by means of the relation
$Z(\omega)=1/\sqrt{\varepsilon(\omega)}$ (which holds, e.g., in
the region of infrared optics). This underlines the fundamental
role of the impedance boundary condition as an alternative to the
consideration of a fluctuating field inside a medium
described by $\varepsilon(\omega)$
in the case of real metals.

We conclude this section by remarking that the obtained expression
(\ref{42}) for the free energy gives the possibility also to find
the thermal Casimir force in configuration of a sphere (spherical
lens) above a plate made of real metals in the
surface impedance approach
\beq F(a)=2\pi R {\cal{F}}_R(a),
\label{45}
\eeq \ni
where $R$ is the sphere radius. The
approximate expression (\ref{45}) is obtained by the application
of the proximity force theorem \cite{6} and has an accuracy around
a fraction of a percent for configurations used in precision
experiments on the measurement of the Casimir force
\cite{7,8,9,10,11,12,14,18}. Thus, the impedance approach provides
the theoretical basis for the measurements of the thermal Casimir
force between real metals to be performed in near future.

\section{Calculation of the Casimir energy in the surface
impedance approach}

First, we apply the obtained general formulas at zero temperature.
In this case Eq.~(\ref{42}) for the free energy transforms to the
double integral representing the Casimir energy between plates
[or, according to Eq.~(\ref{45}), the Casimir force acting between
a sphere and a plate]
    \beq
E(a)=\frac{\hbar}{4\pi^2}
\int_{0}^{\infty}k_{\!\bot}\,d k_{\!\bot} \int_{0}^{\infty}d\xi
\left\{\ln\left[1-r_{\|}^2(\xi, k_{\!\bot})e^{-2aq}\right]\right.
\left. +\ln\left[1-r_{\bot}^2(\xi,
k_{\!\bot})e^{-2aq}\right]\right\}, \label{46} 
\eeq \ni 
where the
reflection coefficients in terms of the impedance are given by
Eq.~(\ref{41}) with the substitution
    \beq q_l\to q=\sqrt{k_{\!\bot}^2+\frac{\xi^2}{c^2}}, \quad Z_l\to Z(i\xi),
    \quad
\xi_l\to\xi. \label{47} \eeq

Let us calculate the quantity (\ref{46}) obtained in the impedance
approach and compare the results with the available data found by
the traditional computations using the Lifshitz formula. For the
purpose of numerical computations, it is convenient to rearrange
Eq.~(\ref{46}) to the form \cite{30}
    \beq
E(a)=\frac{\hbar c}{32\pi^2 a^3} \int_{0}^{\infty}d\zeta
\int_{\zeta}^{\infty}y\,dy \left\{2\ln\left(1-e^{-y}\right)
+\ln\left[1+\frac{X^{\|}(\zeta,y)}{e^{y}-1}\right]
+\ln\left[1+\frac{X^{\bot}(\zeta,y)}{e^{y}-1}\right] \right\},
\label{48} 
\eeq \ni 
where the dimensionless variables $\zeta,\,y$
are defined as \beq \zeta =\frac{\xi}{\omega_c}=\frac{2a\xi}{c},
\qquad y=2qa, \label{49} \eeq \ni and the quantities
$X^{\|,\bot}(\zeta,y)$ are given by
    \beq
X^{\|}(\zeta,y)=\frac{4\zeta yZ}{(y+\zeta Z)^2},
\quad
X^{\bot}(\zeta,y)=\frac{4\zeta yZ}{(\zeta +yZ)^2}, \quad
Z\equiv Z\left(i\frac{c\zeta}{2a}\right).
\label{50}
\eeq
    \ni
The first contribution in the right-hand side of Eq.~(\ref{48})
describes the case of an ideal metal
    \beq E^{(0)}(a)=\frac{\hbar
c}{16\pi^2 a^3} \int_{0}^{\infty}d\zeta \int_{\zeta}^{\infty}y\,dy
\ln\left(1-e^{-y}\right)=-\frac{\pi^2\hbar c}{720 a^3}, \label{51}
\eeq \ni the others are the corrections due to the finite
conductivity.

As was stressed in Sec.~II, with the decrease of temperature the
range of application of the normal skin effect (\ref{1}) reduces
to zero, and at $T=0$ only the anomalous skin effect and infrared
optics occur with the frequency regions given by Eqs.~(\ref{3}),
(\ref{5}), respectively. The transition frequency $\Omega$ between
the two effects can be obtained from the equations
    \beq
\delta_a(\Omega)=\frac{v_F}{\Omega}=\delta_r=\frac{c}{\omega_p},
\label{52}
    \eeq \ni
where, according to Eq.~(\ref{4}),
$\delta_a(\Omega)=C_a/\Omega^{1/3}$. All computations given below
are performed for $Au$ with $\omega_p=1.37\times 10^{16}\,$rad/s
\cite{47} and $v_F=1.4\times 10^6\,$m/s (see, e.g., \cite{48}).
Then from Eq.~(\ref{52}) we obtain the values of both
$C_a=8.8\times 10^{-4}\,\mbox{m}\,\mbox{rad}^{1/3}/\mbox{s}^{1/3}$
and $\Omega=6.36\times 10^{13}\,$rad/s. If to consider $\Omega$
as the characteristic frequency giving the main contribution to
the Casimir effect ($\Omega=\omega_c=c/2a_{tr}$), the transition
separation distance between the two effects turns out to be equal
to $a_{tr}=2.36\,\mu$m. Then it follows that at distances
$\lambda_p<a\ll a_{tr}= 2.36\,\mu$m the impedance of the infrared
optics determines the value of the Casimir energy and force,
whereas at $a\gg a_{tr}=2.36\,\mu$m the impedance of the anomalous
skin effect is applicable ($\lambda_p$=137\,nm is the plasma
wavelength for $Au$). Direct calculations by Eqs.~(\ref{48}),
(\ref{50}) show that the main contribution to the Casimir energy
is given by the narrow frequency interval around the
characteristic frequency $\omega_c$. Thus, the interval
$(0.1\omega_c,10\omega_c)$ contributes 94\% of the total energy in
the wide separation region. What is even more important, the
remainder does not depend on the form of the impedance function
outside of the interval $(0.1\omega_c,10\omega_c)$, to within the
error of about 0.5\%. From this it follows that at each separation
distance between the plates one should, first, determine the
characteristic frequency $\omega_c$ and, second, fix the proper
impedance function. Thereafter the chosen impedance function can
be used at all frequencies when performing the integration in
Eq.~(\ref{48}). At zero temperature this prescription is optional.
At $T\neq 0$, however, it takes on great significance (see
Sec.~VI).

In Fig.~2 the correction factor to the Casimir energy
$E(a)/E^{(0)}(a)$ is plotted which is computed by Eqs.~(\ref{48}),
(\ref{50}) and (\ref{51}) as a function of the separation
distance. The solid line is obtained with the impedance of the
infrared optics (\ref{10}), and the dotted line with the impedance
of the anomalous skin effect (\ref{9}). Both lines are plotted at
all separations $a>\lambda_p$ to make sure that each impedance
function is applicable within its own frequency region and to
follow their applicability at the transition separations around
$a_{tr}$. It must be emphasized that the solid line coincides with
the correction factor to the Casimir energy computed on the basis
of the usual Lifshitz formula in combination with the dielectric
function of the plasma model (this was demonstrated in detail in
Ref.~\cite{30}). Thus, both the impedance approach and the
Lifshitz formula combined with the plasma model, lead to one and
the same result for the Casimir energy at separations
$a>\lambda_p$.

As is seen from Fig.~2, at $\lambda_p<a\ll 2.36\,\mu$m, the
pointed line computed with the impedance of the anomalous skin
effect (which is inapplicable in this region) significantly
underestimates the correction factor due to the finite
conductivity. For example, at $a=0.15\,\mu$m the values of the
correction factors, given by the solid and dotted lines, are 0.623
and 0.851, respectively, i.e., the error introduced by the use of
the impedance of the anomalous skin effect is almost 37\%. At a
separation $a=0.5\,\mu$m this error is more than 9\%, and
decreases with increasing separation. Notice that the computations
on the basis of the usual Lifshitz formula and optical tabulated
data for the complex refraction index [which are used to obtain
$\varepsilon(i\xi)$ through the dispersion relation] also
practically coincide with those given by the surface impedance in
the region of the infrared optics (solid line in Fig.~2). Thus, at
the separations of $0.2\,\mu$m, $0.5\,\mu$m and $3\,\mu$m the
correction factor obtained by the tabulated data and Lifshitz
formula is equal to 0.69, 0.85, and 0.97, respectively
\cite{45,47}, whereas in the impedance approach it takes the
values 0.689, 0.849, and 0.972.

At larger separations ($a\gg a_{tr}=2.36\,\mu$m) the impedance
function of the anomalous skin effect should be used to compute
the Casimir energy (dotted line in Fig.~2). As is seen from that
figure, at these separations the impedance of the infrared optics
overestimates the role of the finite conductivity corrections to
the Casimir energy. This overestimation is, however, to within a
fraction of a percent. In the transition region $a=2-2.5\,\mu$m
the results given by both impedance functions are in agreement
bringing the discrepancies of about 1\% only. This leads us to the
conclusion that at zero temperature both impedance functions work
well in their respective application regions. In the transition
region each of them can be applied and the results are in
agreement within an error of 1\%. It is seen also that ``much
less'' or ``much larger'' in the above inequalities in fact mean
two or three times less (larger).

If to speak about the region of infrared optics, the Lifshitz
formula in combination with the plasma model or optical tabulated
data for the complex refraction index leads to the same results as
the impedance approach. It gives rather good results even in the
region of the anomalous skin effect, where, strictly speaking, the
description in terms of $\varepsilon$ is not applicable (see
Sec.~II). The feasibility of the Lifshitz formula is explained by
the fact that at zero temperature the normal skin effect is
practically absent and the problems connected with the heating of
a metal due to the real electric current are not relevant. As a
result, both the impedance approach and the
usual
Lifshitz formula are
applicable. At nonzero temperature, however, the surface impedance
approach acquires a new meaning and solves the problems formulated
in Introduction (see the next section).

\section{Calculation of the Casimir free energy in the surface
impedance approach}

Here we calculate the Casimir free energy for the configuration of
two parallel plates made of $Au$ at a temperature $T$ at thermal
equilibrium. The starting point is Eq.~(\ref{42}) where the
reflection coefficients are expressed in terms of the surface
impedance by Eq.~(\ref{41}). Introducing the dimensionless
variables by analogy with Eq.~(\ref{49}), we transform
Eq.~(\ref{42}) to a form convenient for numerical computations:
    \bes
&&
{\cal{F}}_R={\cal{F}}_R(a,T)=\frac{k_BT}{8\pi a^2}
\sum\limits_{l=0}^{\infty}{\vphantom{\sum}}^{\prime}
\int_{\zeta_l}^{\infty}y\,dy
\left\{2\ln\left(1-e^{-y}\right)\right.
\nn \\
&& \phantom{aaa}\left.
+\ln\left[1+\frac{X^{\|}(\zeta_l,y)}{e^{y}-1}\right]
+\ln\left[1+\frac{X^{\bot}(\zeta_l,y)}{e^{y}-1}\right] \right\},
\label{53} \ees \ni where $X^{\|},\,X^{\bot}$ are given by
Eq.~(\ref{50}) with the change $\zeta\to\zeta_l$, $Z\to
Z_l=Z(ic\zeta_l/2a)$. Notice that the first contribution in the
right-hand side of Eq.~(\ref{53}) presents the Casimir free energy
for the ideal metal \cite{6}
    \bes &&
{\cal{F}}_R^I(a,T)=\frac{k_BT}{4\pi a^2}
\sum\limits_{l=0}^{\infty}{\vphantom{\sum}}^{\prime}
\int_{\zeta_l}^{\infty}y\,dy \ln\left(1-e^{-y}\right)
\label{54} \\
&&
\phantom{a}
=E^{(0)}(a)\left\{1+
\frac{45}{\pi^3}
\sum\limits_{l=1}^{\infty}
\left[\left(\frac{T}{T_{eff}}\right)^3\frac{1}{l^3}
\coth\left(\pi l\frac{T_{eff}}{T}\right)\right.\right.
\nn \\
\phantom{aaaa}
&&
\left.\left.
+\pi \left(\frac{T}{T_{eff}}\right)^2\frac{1}{l^2}
\sinh^{-2}\left(\pi l\frac{T_{eff}}{T}\right)\right]
-\left(\frac{T}{T_{eff}}\right)^4\right\},
\nn
\ees
\ni
where $E^{(0)}(a)$ is defined in Eq.~(\ref{51}) and the effective
temperature $k_BT_{eff}=\hbar\omega_c=\hbar c/(2a)$.

First of all, let us demonstrate that in the impedance approach
there is no problem with the contribution of the zero Matsubara
frequency which was the subject of much recent controversy (see
Introduction). We start from the lowest characteristic frequencies
where the impedance of the normal skin effect, given by
Eq.~(\ref{8}), is applicable. Substituting it into Eq.~(\ref{41})
and putting $\xi_l=0$, one obtains \beq r_{\|}^2(0,k_{\!\bot})=
r_{\bot}^2(0,k_{\!\bot})=1, \label{55} \eeq \ni i.e., the same
result as for an ideal metal.
Namely, this becomes clear since the quantities $X^{\|,\bot}$,
defined in Eq.~(\ref{50}) and given for the impedance of the
normal skin effect as
    \beq
    X^{\|}(0,y)=X^{\bot}(0,y)=0,
\label{56}
    \eeq \ni
when inserted into Eq.~(\ref{53}) obviously lead to the same
zero-frequency contribution as it holds for an ideal metal.
It should be stressed that all functions
$r_{\|,\bot}^2(\xi,k_{\!\bot})$ and $X^{\|,\bot}(\zeta,y)$ are
continuous functions of two variables including the point (0,0).
Thus, the case of an ideal metal is achieved as a limiting case of
a real metal with increase of the conductivity when the real metal
is described in the framework of the impedance approach. Remind
that this is not the case when the real metal is described by the
Drude dielectric function (\ref{2}) and the Lifshitz formula for
dielectrics is used to calculate the Casimir free energy [approach
(a) from Introduction]. In fact, if doing so it follows from
Eq.~(\ref{44})
\beq r_{\|,L}^2(0,k_{\!\bot})=1,\quad
r_{\bot,L}^2(0,k_{\!\bot})=0,
\label{57}
\eeq \ni
and there is a
break of continuity between the properties of real metals and of
ideal metal \cite{28}.

At higher characteristic frequencies the anomalous skin effect
holds with an impedance function as from Eqs.~(\ref{4}),
(\ref{9}). If to extend this function to all frequencies (to zero
Matsubara frequency in that case) we ensure that both the
Eqs.~(\ref{55}) and (\ref{56}) are valid once again. As a result,
in both regions of the normal and the anomalous skin effect the
thermal corrections to the Casimir free energy and force for real
metals are very close to those for an ideal metal. As one would
expect, at large separations (characteristic for the anomalous
and especially for the normal skin effect) all metals behave
like an ideal one [this is, however, not the case in the framework
of the approach (a)].

If the characteristic frequencies increase further, the infrared
optics with an impedance function of Eq.~(\ref{10}) takes place.
The extension of it to zero Matsubara frequency leads to
    \beq
r_{\|}^2(0,k_{\!\bot})=1,\quad
r_{\bot}^2(0,k_{\!\bot})=
\left(\frac{\omega_p-ck_{\!\bot}}{\omega_p+ck_{\!\bot}}\right)^2.
\label{58}
\eeq
    \ni
In this case there occurs a dependence of the perpendicular
reflection coefficient at zero frequency on the properties of the
real metal through the value of the plasma frequency. This is
reasonable, because the real properties of a metal are most
pronounced at small separations characteristic for the infrared
optics. In the limit $\omega_p\to\infty$ the result for an ideal
metal is reproduced from Eq.~(\ref{58}).

Before performing the computations, it must be emphasized that the
surface impedance approach is in perfect agreement with
thermodynamics. In the impedance approach the entropy, defined as
    \beq
S(a,T)=-\frac{\partial{\cal{F}}_R(a,T)}{\partial T}, \label{59}
\eeq
    \ni
is positive and equal to zero at zero temperature in accordance
with the Nernst heat theorem [remind that this is not the case in
the approaches (a) and (c)]. The validity of the Nernst heat
theorem in the impedance approach can be demonstrated in the
regions of both the infrared optics and the anomalous skin effect
(as noted above, the region of the normal skin effect dies out
with decreasing temperature). According to the results of
Ref.~\cite{30}, in the region of infrared optics the Lifshitz
formula combined with the plasma model leads to exactly the same
perturbation results for the Casimir free energy and force as the
impedance approach. At $T\ll T_{eff}$ the free energy is given by
\cite{31}
    \beq 
{\cal{F}}_R(a,T)=E(a)-\frac{\hbar
c\zeta(3)}{16\pi a^3} \left[\left(1+2\frac{\delta_r}{a}\right)
\left(\frac{T}{T_{eff}}\right)^3
-\frac{\pi^3}{45\zeta(3)}\left(1+4\frac{\delta_r}{a}\right)
\left(\frac{T}{T_{eff}}\right)^4\right], 
\label{60} \eeq \ni 
where
$E(a)$ is the Casimir energy at $T=0$ defined in Eq.~(\ref{46}).
After the substitution in Eq.~(\ref{59}), this leads to the simple
expression for the Casimir entropy 
\beq
S(a,T)=\frac{3k_B\zeta(3)}{8\pi
a^2}\left(\frac{T}{T_{eff}}\right)^2 \left\{
\vphantom{\left[\frac{8\pi^3}{135\zeta(3)}\frac{T}{T_{eff}}\right]}
1-\frac{4\pi^3}{135\zeta(3)} \frac{T}{T_{eff}}
+2\frac{\delta_r}{a}\left[
1-\frac{8\pi^3}{135\zeta(3)}
\frac{T}{T_{eff}}\right]\right\},
\label{61}
\eeq
\ni
which is positive and equal to zero at zero temperature.

The impedance approach in the region of the anomalous skin effect
was used in the recent Ref.~\cite{32}. The asymptotic of the
entropy at very low temperatures, obtained in Ref.~\cite{32},
demonstrates that it is positive and has zero value at zero
temperature in accordance with the requirements of thermodynamics.

By the way of example, here we perform the numerical computations
of the relative thermal correction to the Casimir free energy
defined as $[{\cal{F}}_R(a,T)-E(a)]/E(a)$. This quantity has also
the meaning of the relative thermal correction to the Casimir
force in the configuration of a sphere (spherical lens) above a
plate used in precision experiments on the Casimir effect. If the
characteristic frequency $\omega_c$ belongs to the region of the
normal skin effect, the results practically coincide with those
obtained for an ideal metal \cite{30}, and the free energy is
given by Eq.~(\ref{54}). If the characteristic frequency belongs
to the regions of the anomalous skin effect or infrared optics,
the computational results for the relative thermal correction are
obtained by Eqs.~(\ref{48}), (\ref{53}) and presented in Fig.~3.
The solid lines are computed with the impedance of the infrared
optics (\ref{10}), and the dotted lines with the impedance of the
anomalous skin effect (\ref{9}). All computations are performed
for $Au$ at two temperatures $T=300\,$K and $T=70\,$K with
numerical parameters as listed in Sec.~V. Both pairs of lines are
plotted at all separations $a>\lambda_p$ for a better
visualization of the application range of each impedance function.

Remind that at separations between the plates $\lambda_p<a<a_{tr}$,
where $a_{tr}=2.36\,\mu$m does not depend on the temperature, the
impedance of the infrared optics is applicable, and at separations
$a>a_{tr}$ the impedance of the anomalous skin effect should be
used. It is seen from Fig.~3 that at small separations the use of
the impedance function of the anomalous skin effect significantly
overestimates the value of the thermal correction. Thus, at
$a=0.15\,\mu$m the values of the relative thermal corrections
given by the dotted and solid lines are $1.55\times 10^{-2}$ and
$1.82\times 10^{-4}$, respectively, at $T=300\,$K, and $4.85\times
10^{-3}$ and $2.76\times 10^{-6}$, respectively, at $T=70\,$K.
What this means is the thermal correction, predicted by the
impedance of the anomalous skin effect in the region of the
infrared optics, where this impedance is not applicable, is in 85
times greater at $T=300\,$K and in 1757 times greater at $T=70\,$K
than the correct values.

At separations $a>a_{tr}=2.36\,\mu$m the dotted lines present the
correct dependence of the thermal correction on the separation
distance. The difference between the free energies computed by the
use of two impedance functions is, however, to within a fraction
of a percent. In the transition region the results, given by the
impedance function of the infrared optics and anomalous skin
effect, are in agreement with a sufficient accuracy. For example,
at $a=2.5\,\mu$m the ratio of the relative thermal corrections
obtained by the use of different impedance functions is 1.05 at
$T=300\,$K and 2.19 at $T=70\,$K. As a result, the discrepancies
in the values of the free energy are about 1.2\% ($T=300\,$K) and
0.7\% ($T=70\,$K).

Our results for the thermal correction to the Casimir free energy
are in disagreement with the conclusion of Ref.~\cite{32} about
the existence of large thermal corrections at low temperature made
in the framework of the impedance approach. As correctly argued in
Ref.~\cite{32}, the description of metals with the impedance in
the region of the anomalous skin effect is more appropriate than
with the dielectric permittivity. However, the conclusion about
the existence of large thermal corrections at separations
100\,nm--500\,nm at $T\leq 70\,$K made in that paper is in error.
To obtain this conclusion, the impedance function of the anomalous
skin effect was applied in Ref.~\cite{32} at separations much less
than $a_{tr}=2.36\,\mu$m, i.e.,  in the separation range of the
infrared optics. This was explained by the fact that at
temperatures $T\leq 70\,$K the inequality $l\gg\delta_r$ holds
which, from the standpoint of Ref.~\cite{32}, guarantees the
applicability of the impedance of the anomalous skin effect. In
actual truth, this inequality is not sufficient. In fact, one
additional inequality, $\delta_a(\omega)\ll v_F/\omega$, must be
fulfilled in order that the anomalous skin effect holds [see
Eq.~(\ref{3}) and Ref.~\cite{37}]. Because of this, the frequency
$\Omega$ [see definition in Eq.~(\ref{52})], considered in
Ref.~\cite{32} as the characteristic frequency of the anomalous
skin effect, is actually the transition frequency to the region of
infrared optics. As a result, all computations performed in
Ref.~\cite{32} correspond to the dotted line at $T=70\,$K of our
Fig.~3 at separations $a<a_{tr}=2.36\,\mu$m. According to our
computations, in this separation range the dotted line at
$T=70\,$K overestimates the value of the thermal correction by a
factor of 2000, whereas the correct results are given by the solid
lines obtained by the use of the impedance function of infrared
optics. Remind that the characteristic frequencies, corresponding
to the separations 100\,nm--500\,nm, fall into the interval
$\omega_c=(0.3-1.5)\times 10^{15}\,$rad/s$\gg\Omega$, i.e., belong
to the region of the infrared optics (see Sec.~II).

At the end of this section, we would like to stress that in the
sums, like in Eqs.~(\ref{42}), (\ref{43}) and (\ref{53}), the form
of the impedance at the characteristic frequencies must be
substituted and extended to all other frequencies. At zero
temperature, as was shown in Sec.~V, the frequency region
$[0,1\omega_c,10\omega_c]$, where the characteristic frequency is
$\omega_c=c/(2a)$, gives most of the contributions to the result.
Calculations show that at nonzero temperature the Matsubara
frequencies from $\xi_0$ to $\xi_N\approx 10\omega_c$ give the
dominant contribution. For example, at $a=0.15\,\mu$m
($\omega_c=10^{15}\,$rad/s), $T=300\,$K the first 41  Matsubara
frequencies determine the total result. Here $\xi_1=2.5\times
10^{14}\,$rad/s and $\xi_{40}=10^{16}\,$rad/s. All
nonzero Matsubara frequencies belong
to the region of infrared optics. With a decrease of the
temperature some of the Matsubara frequencies may fall within the
frequency region of the anomalous skin effect (at $T=70\,$K, for
instance, the first Matsubara frequency $\xi_1=5.75\times
10^{13}\,$rad/s$<\Omega=6.36\times 10^{13}\,$rad/s for $Au$). At
small separations, however, the differences in the contributions
of several first Matsubara frequencies computed by the use of
different impedance functions are negligible. At $T=0$ any
extension of the impedance function outside of the above interval
leads to approximately one and the same value of the Casimir
energy (in the integral one point $\xi=0$ is of no significance).
At $T\neq 0$, however, the contribution of the zero Matsubara
frequency $\xi_0=0$ becomes dominant at large separations (high
temperatures), and at room temperature, for instance, it
determines the total value of the free energy at $a\geq 5\,\mu$m.

The basic challenge is whether the actual reflection properties of
plate materials at very low, quasi-static, frequencies are
responsible for the Casimir force in the high-temperature limit.
The point is that materials are at hand (e.g., indium tin oxide)
which are good conductors at quasi-static frequencies but
transparent to visible and near infrared light. If to consider a
pair of plates made of indium tin oxide (ITO) at a separation
$a=5\,\mu$m, and the other pair of plates at the same separation
made of $Au$, one runs into difficulties. If the actual
low-frequency reflection properties should be substituted into the
zero-frequency term, the impedance of the normal skin effect from
Eq.~(\ref{8}) must be used. As a result, the thermal Casimir force
at $a=5\,\mu$m will be equal for both pairs of plates (and
practically the same as for an ideal metal). This is in
contradiction with the physical intuition as around the
characteristic frequency $\omega_c=3\times 10^{13}\,$rad/s
(computed at the separation $5\,\mu$m) ITO is a poor reflector. A
better physical result would be obtained if one extends the
characteristic impedance at $\omega_c$ (of the anomalous skin
effect for $Au$ and of the infrared optics for ITO) to zero
Matsubara frequency. If this is done, the zero-frequency term for
$Au$ plates will be the same as for an ideal metal in accordance
with Eq.~(\ref{55}). For ITO plates the zero-frequency term will
contain the value of $\omega_p^{ITO}$ according to Eq.~(\ref{58}).
Taking into account the large value of the penetration depth for
ITO, the magnitude of the Casimir force between the ITO plates
will be less than between the plates made of $Au$, as the
intuition suggests. In fact, the question on whether the values of
the Casimir force for the two above pairs of plates at separation
5$\,\mu$m are equal or different, can be answered experimentally
using the measurement scheme suggested recently in Ref.~\cite{49}.
We expect that the experimental result will be in accordance with
the suggestion of the physical intuition (note that this example
with two pairs of plates was used with another aim in
Ref.~\cite{31}).

\section{Conclusions and discussion}

In the foregoing we have presented the surface impedance approach
to the theory of the Casimir effect at both zero and nonzero
temperature. In the impedance approach the effective boundary
condition is imposed taking the real properties of the metal into
account. Previously this approach was considered as nothing more
than a useful approximation to the more complete Lifshitz theory
using the concept of a fluctuating electromagnetic field both
outside and inside the boundary of the bodies. Our conclusion is
that the
standard
concept of a fluctuating field inside a metal,
described by the dielectric permittivity depending only on
frequency, in the region
where a
real current may arise, can not serve as an adequate model for the
zero-point oscillations and thermal photons. It follows from the
fact that the vacuum oscillations and the thermal photons in
equilibrium under no circumstances can lead to a heating of the
metal. If this fact is overlooked, contradictions with the
thermodynamics arise when one substitutes into the Lifshitz
formula for the Casimir free energy and force the Drude dielectric
function which takes into account the volume relaxation and,
consequently, the Joule heating.
This situation reflects the nontrivial character of quantum
fluctuations in nonhomogeneous case involving both vacuum and
real metals, containing conduction electrons, in different
spatial regions. In fact, in such cases the quantized
electromagnetic field at nonzero temperature may not be
represented in terms of (quasi)-particles \cite{31}.
As a result, the concept of a fluctuating field becomes
not so transparent as in nonlossy dielectric media.

In the light of this conclusion, the surface impedance approach
takes on fundamental importance as
(at present)
the only self-consistent
description of the Casimir effect between real metals. It does not
need any prescription for the zero-frequency contribution to the
Casimir energy and force. In all cases the correct expressions for
the values of both reflection coefficients with two different
polarizations at zero frequency are deduced from the general
theoretical framework using the explicit form of the impedance
function (see Sec.~VI). Thus, a long discussion in the recent
literature concerning the most adequate modification of the
zero-frequency term of the Lifshitz formula
\cite{21,23,24,25,26,27,28,31,32,33} can be finalized.

The surface impedance approach solves the puzzle with the
violation of the Nernst heat theorem and with negative values of
entropy which appears when one substitutes the Drude dielectric
function into the
usual
Lifshitz formula. In the impedance approach the
entropy is in all cases nonnegative and takes zero value at zero
temperature. Thus, the general formulas given by Eqs.~(\ref{41}),
(\ref{42}), (\ref{43}), (\ref{46}), (\ref{48}), and (\ref{53}) lay
the theoretical basis for the calculation of the thermal Casimir
effect with respect to the needs of future precision experiments.
The computations performed in Secs.~V,\,VI are in good agreement
with the previous results obtained by the use of the optical
tabulated data and the plasma model.

The obtained results allow to remove the doubts that something is
wrong with the Lifshitz formula \cite{32}. In fact, the above
formulas in the framework of the impedance approach coincide with
the Lifshitz formula. The only difference is that for real metals
in the frequency regions, where the electromagnetic oscillations
initiate a real current or where the space dispersion is
essential, one must express the reflection coefficients not in
terms of the dielectric permittivity but in terms of the surface
impedance. The usual Lifshitz formula, formulated in terms of the
dielectric permittivity, preserves, however, major importance not
only in applications to dielectrics but also in the theory of the
non-retarded van der Waals forces between metallic surfaces. As
was mentioned in Sec.~II, the surface impedance approach is
applicable with the proviso that $\omega_c<0.1\omega_p$, i.e., the
separation distances between the test bodies must satisfy the
condition $a>5\lambda_p/(2\pi)\approx\lambda_p$. In essence, the
frequency region $\omega>0.1\omega_p$ is a subject of the optics
of real metals near the plasma frequency \cite{50}. At separations
$a<\lambda_p$ between the test bodies the temperature effects are
negligible. This is a region of the ultraviolet transparency where
metals can be described on the same basis as dielectrics. The most
adequate approach to the theory of the van der Waals forces at so
small separations is given by the hydrodynamical description of an
inhomogeneous electron gas \cite{51}. This is a more general
approach if compared with the Lifshitz theory because it does not
start with a model description of a metal in terms of the bulk
dielectric permittivity. In the local limit, however, when the
spatial dispersion is absent, the hydrodynamical approach leads to
the usual Lifshitz formula at zero temperature \cite{51}. As a
consequence, the usual Lifshitz formula is well adapted for the
calculation of the van der Waals forces between real metals at
separations $a<\lambda_p$ if ${\e}(i\xi)$ being obtained by the
use of optical tabulated data for the complex refraction index
(the extension of the avalible tabulated data into the region of
small frequencies makes almost no effect on the computational
results).

In conclusion it may be said that the Lifshitz formula in
combination with the impedance approach gives a solid foundation
for the investigation of thermal effects onto the Casimir force.
This approach does not lead to contradictions and can be used as
the theoretical basis for the needs of future experiments.

\section*{Acknowledgments}

G.L.K.  is greatly indebted to I.E.Dzyaloshinskii for attracting her
attention to the fundamental importance of the concept of surface
impedance.
G.L.K. and V.M.M. are grateful to the Center of Theoretical Studies and
the Institute for Theoretical
Physics, Leipzig University for kind hospitality. Their work was
supported by the Saxonian Ministry of Science and Fine Arts (Germany)
and by CNPq (Brazil).

\newpage
\widetext
\begin{figure}[h]
\vspace*{-4cm}
\epsfxsize=20cm\centerline{\epsffile{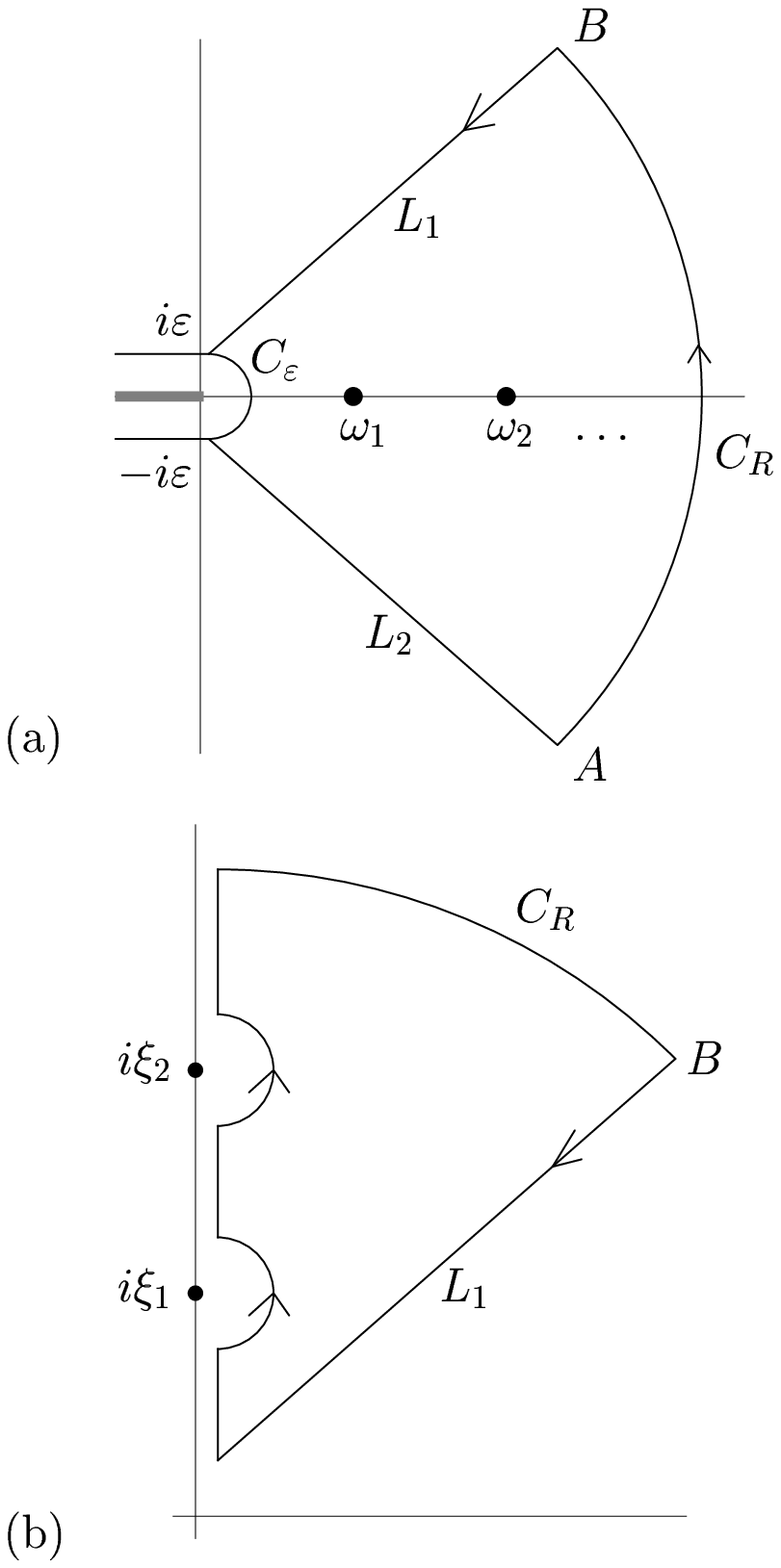}}
\vspace*{-5cm}
\caption{Integration paths $C_1$ (a) and $C_2$ (b) in the plane
of complex frequency.
The Matsubara frequencies are $\xi_l$ and photon eigenfrequencies
are $\omega_n$.
}
\end{figure}
\newpage
\widetext
\begin{figure}[h]
\vspace*{-7cm}
\epsfxsize=20cm\centerline{\epsffile{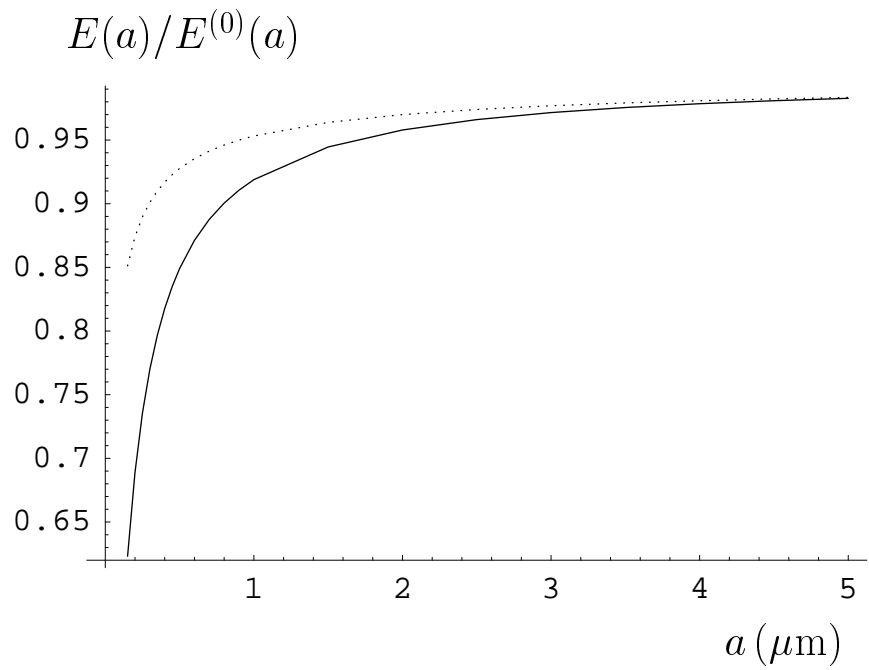}}
\vspace*{-8cm}
\caption{Correction factor to the Casimir energy between two $Au$ plates
at zero temperature computed by the use of the impedance of infrared
optics (solid line) and of anomalous skin effect (dotted line)
versus surface separation.
}
\end{figure}
\newpage
\widetext
\begin{figure}[h]
\vspace*{-3cm}
\epsfxsize=20cm\centerline{\epsffile{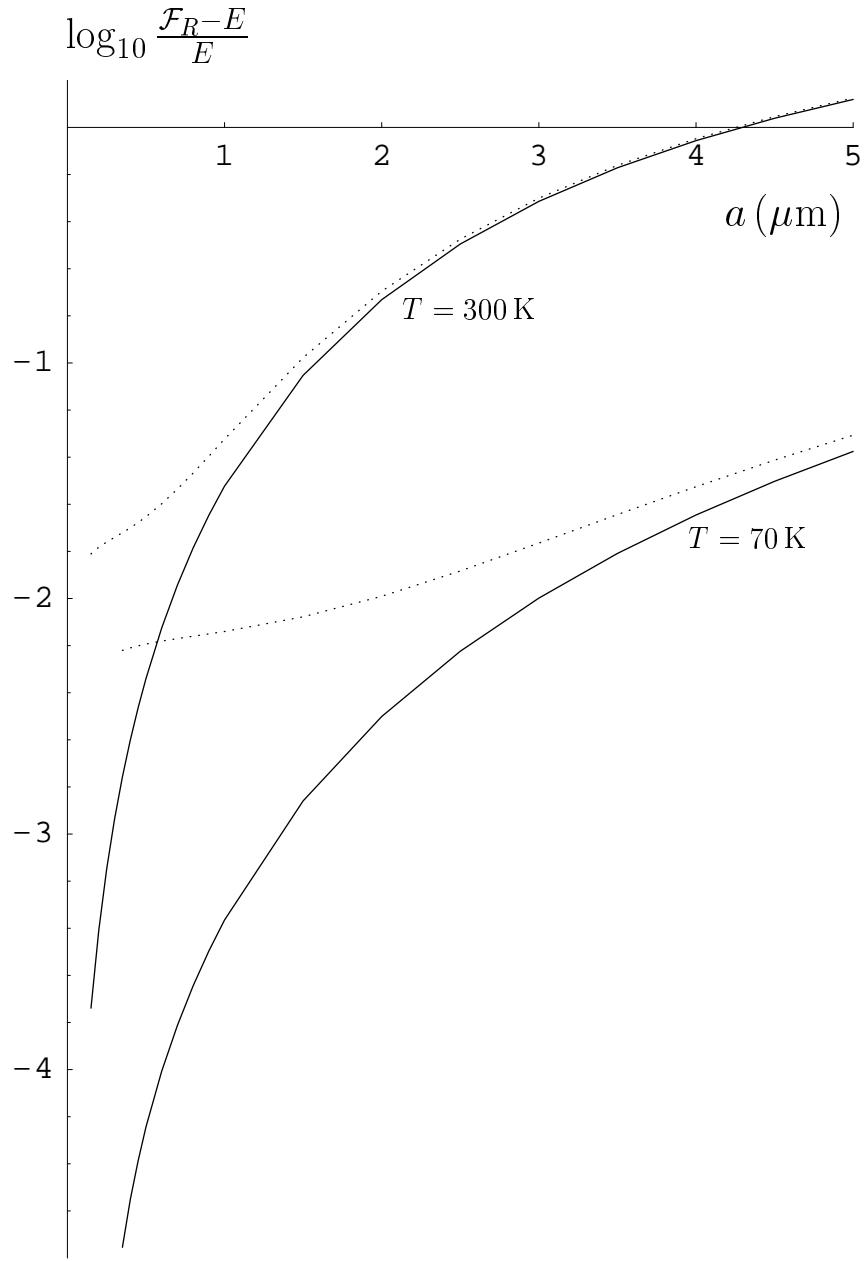}}
\vspace*{-5cm}
\caption{Relative thermal  correction to the
Casimir free energy between two $Au$ plates
computed by the use of the impedance of infrared
optics (solid lines) and of anomalous skin effect (dotted lines)
versus surface separation at two different temperatures.
}
\end{figure}
\end{document}